# On the Quantitively Characterization of Intermittent Power Sources Uncertainty

**Zongjie Wang [1,\*], Zhizhong Guo [1, 2]**

[1]  School of Electrical Engineering and Automation, Harbin Institute of Technology, Harbin 150001, China
     fylx-001@163.com (W.Z.); zhizhonggzz@sina.com (G.Z.)

[2]  Harbin Institute of Technology at Zhangjiakou ITRIZ, Zhangjiakou 075400, China; zhizhonggzz@sina.com
     (G.Z.)

\*  Correspondence: fylx-001@163.com; Tel.: +86-189-1029-4513

**Abstract:** This paper designs a statistical quantification towards the intermittent power uncertainty in power systems. A negative-exponential forecast uncertainty function is constructed to represent the relationship between the statistics of forecast error of a single intermittent power source and time advance. Subsequently, other kinds of statistical functions are proposed to characterize the statistical uncertainty of multiple intermittent power sources and all power sources, namely the sum statistical functions, the equivalent statistical functions, and the contour statistical functions. Based on a large amount of historical observations, these functions are employed to statistically quantify the forecast uncertainty of a single intermittent power source, multiple intermittent power sources as well as all power sources. Historical data sampled from real wind farms and solar sites demonstrates the effectiveness of the proposed method.

## 1. Introduction

Intermittent Power Sources (IPS) such as wind and solar generators have been extensively integrated into power systems [1-3]. However, the IPS power generation is highly unpredictable due to their uncertainties such as intermittence, randomness and volatilities [4-6]. The overall uncertainty level of a power system depends on the uncertainty level of each single IPS and the proportion of the IPS power generation [7-11]. A report from National Renewable Energy Lab (NREL) predicted that the renewable electricity generation is more than adequate to supply 80% of total U.S. electricity generation in 2050 [12]. Such a large amount of unpredictable power generation may jeopardize the securities and stabilities of power systems. Indeed, the uncertainty brought by IPS has been widely acknowledged as one of the most significant challenges to the real-time power balancing [13-15].

The essence of uncertainty brought by IPS is the unpredictability due to time advances, which can be represented by forecast error. Several algorithms for day-ahead forecast of IPS power generation have been developed [16-18]. However, to the best of our knowledge, obtaining satisfactory forecast accuracy for IPS power generation is still an open problem. The intrinsic reason is that the traditional time advance of 24 hours is too far to obtain an accurate forecast for IPS power generations. To ameliorate the status quo, the most direct method is to shorten the time advance of power forecast. However, this also reduces the time reserved for dispatch and market operations. To come up with the best time advance for the power forecast of IPS, a quantitative and statistical evaluation of the overall uncertainty level of all power sources is highly desirable. Namely, we are interested in the temporal characteristics of the statistical forecast uncertainty.

There are many publications regarding the statistics of power generation uncertainty [19-26]. Paper [19] employed the maximum entropy spectrum estimation method to study the temporal and spatial distribution characteristics of wind power periodicity. Paper [20] presented an advanced statistical method for wind power forecast based on artificial intelligence technologies. Paper [21] studied the statistical regularity of wind power fluctuation in different spatial-temporal scales from two aspects of smoothness and correlation. In particular, statistical studies on the relationship between the forecast error of IPS generation and the time advance of look-ahead dispatch have been presented in [23-26]. Take papers [24, 25] as an example, it is shown that as the time advance decreases, the forecast error decays slowly at first, but then rapidly diminishes after a particular point. Besides the noise in each single observation, the behavior of an uncertain variable yields to certain statistics that implicate the latent laws thereof. Regarding the forecast error of IPS generation, our objective is to minimize the effect of noise by taking a large number of historical observations and extract its statistical characteristics, which can be approximately characterized by the statistical functions of forecast error designed in this paper.

In this paper, we employ a family of negative-exponential functions to represent the temporal characteristics of statistical forecast uncertainty. Specifically, we define statistical functions of forecast error of a single IPS, the sum statistical function of multiple IPS as well as all power sources that can also be represented by the equivalent statistical function with a variant scaling factor, and the contour statistical function to portray a closure of the sum statistical function. We compare these functions with a large amount of real-data from papers [25, 26] and present an estimation method of the parameters of these statistical functions. These statistical functions can succinctly extract the temporal trend of statistical forecast uncertainty level of all power sources.

The remainder of this paper is organized as follows. We first propose a statistical function of forecast error in Section 2. The uncertainty of IPS with statistical functions is quantified in Section 3. Subsequently, the uncertainty of all power sources with statistical functions is further quantified in Section 4. The simulation results are demonstrated in Section 5. In Section 6, conclusions are drawn and practical applications are finally discussed.

## 2. Statistical Forecast Uncertainty

### 2.1. Statistical Functions of Forecast Error

In this subsection, we first define the Statistical Function of Forecast Error (SFFE) of IPS to characterize the statistical forecast uncertainty of a single IPS with respect to the time advance $t$, represented by $\alpha(t)$ and is shown as follows

$$\alpha(t) \triangleq A(1 - e^{-\frac{t}{\tau}}), \ (t \in [0, +\infty)) \,. \tag{1}$$

Herein, the amplitude $A \in (0, +\infty)$ and the time coefficient $\tau \in (0, +\infty)$ satisfy the following boundary condition

$$\begin{cases} A = \alpha(+\infty) \\ \tau = \dfrac{A}{\alpha^{(1)}(0)} = \dfrac{\alpha(+\infty)}{\alpha^{(1)}(0)} \end{cases} . \tag{2}$$

To better demonstrate the influence of the statistical forecast uncertainty with respect to the time advance $t$, we normalize the $\alpha(t)$ in (1) and define the following Normalized Statistical Function of Forecast Error (NSFFE) that is represented by $\lambda(t)$ as follows

$$\lambda(t) \triangleq \frac{\alpha(t)}{A} = 1 - e^{-\frac{t}{\tau}} \,. \tag{3}$$

Next, we establish some important properties for the SFFE and the NSFFE, i.e., the $\alpha$ function and the $\lambda$ function.

**Proposition 1.** *The NSFFE ($\lambda$ function) yields the following conservation principle*

$$\tau \lambda^{(i)} + \lambda^{(i-1)} = \begin{cases} 1, (i = 1) \\ 0, (i \geq 2) \end{cases}, \tag{4}$$

*where $\lambda^{(i)}$ is the $i^{th}$ -order derivative of the $\lambda$ function.*

**Proof.** The first order derivative of the $\lambda$ function is

$$\lambda^{(1)}(t) = \frac{1}{\tau} e^{-\frac{t}{\tau}}, \tag{5}$$

and the following equation holds

$$\tau\lambda^{(1)}(t) + \lambda(t) = 1 . \tag{6}$$

Note that equation (6) holds for $\forall t \in [0, +\infty)$. Hence by continuously differentiating the equation above we have

$$\tau\lambda^{(i)}(t) + \lambda^{(i-1)}(t) = 0, (i \geq 2) . \tag{7}$$

This ends the proof. □

**Corollary 1.** *The SFFE ($\alpha$ function) yields the following conservation principle*

$$\frac{\alpha^{(i)}(t)}{\alpha^{(1)}(0)} + \frac{\alpha^{(i-1)}(t)}{A} \equiv \begin{cases} 1, (i=1) \\ 0, (i \geq 2) \end{cases} . \tag{8}$$

**Proof of Corollary 1.** Since the following equation holds for all $i$

$$\lambda^{(i)}(t) = \frac{\alpha^{(i)}(t)}{A} , \tag{9}$$

by substituting the equation above into equation (4) we have

$$\tau\frac{\alpha^{(i)}(t)}{A} + \frac{\alpha^{(i-1)}(t)}{A} \equiv \begin{cases} 1, (i=1) \\ 0, (i \geq 2) \end{cases} . \tag{10}$$

Note that

$$\frac{\tau}{A} = \frac{1}{\alpha^{(1)}(0)} , \tag{11}$$

therefore equation (8) holds. This ends the proof. □

### 2.2. Sum Statistical Functions

In order to characterize the sum statistical forecast uncertainty of all IPS, we define the Sum Statistical Function (SSF) with respect to the time advance $t$, represented by $\alpha_{\Sigma}(t)$ and is shown as follows

$$\alpha_{\Sigma}(t) \triangleq \sum_{i=1}^{n} \alpha_i(t) . \tag{12}$$

where $n$ represents the quantity of $\alpha$ functions.

Let

$$A_{\Sigma} \triangleq \sum_{i=1}^{n} A_i , \tag{13}$$

and

$$\rho_i \triangleq \frac{A_i}{A_{\Sigma}}, (i = 1, 2, ..., n) , \tag{14}$$

there is

$$\alpha_{\Sigma}(t) = A_{\Sigma} \sum_{i=1}^{n} \rho_i (1 - e^{-\frac{t}{\tau_i}}) . \tag{15}$$

Similarly, we normalize the $\alpha_{\Sigma}(t)$ in (12) and define the following Normalized Sum Statistical Function (NSSF) that is represented by $\lambda_{\Sigma}(t)$ as follows

$$\lambda_{\Sigma}(t) \triangleq \frac{\alpha_{\Sigma}(t)}{A_{\Sigma}} = \sum_{i=1}^{n} \rho_i (1 - e^{-\frac{t}{\tau_i}}) . \tag{16}$$

Consequently, equation (15) can be transformed into

$$\alpha_{\Sigma}(t) = A_{\Sigma}\lambda_{\Sigma}(t) . \tag{17}$$

### 2.3 Equivalent Statistical Functions

The $\alpha_{\Sigma}(t)$ in (12) is the sum of multiple negative-exponential SFFE. For simplicity, in this subsection, we denote the SSF by a single negative exponential function with a variant time coefficient $\tau(t)$. Such a function is referred to as the Equivalent Statistical Function (ESF), represented by $\alpha(t, \tau(t))$ and is shown as follows

$$\alpha(t, \tau(t)) \triangleq A_\Sigma (1 - e^{-\frac{t}{\tau(t)}}),\tag{18}$$

where the time coefficient $\tau(t)$ is a function of the time advance $t$.

For $\forall t \in [0, +\infty)$ and $n$ given SFFE ($\alpha$ functions) there exists a $\tau(t)$ such that the following equation holds

$$\alpha(t, \tau(t)) = \alpha_\Sigma(t).\tag{19}$$

Similarly, we normalize the $\alpha(t, \tau(t))$ in (18) and define the following Normalized Equivalent Statistical Function (NESF) that is represented by $\lambda(t, \tau(t))$ as follows

$$\lambda(t, \tau(t)) = 1 - e^{-\frac{t}{\tau(t)}}.\tag{20}$$

Obviously,

$$\lambda(t, \tau(t)) = \lambda_\Sigma(t).\tag{21}$$

**Proposition 2.** *Given $n$ NSFFE ($\lambda$ functions), the time coefficient function $\tau(t)$ defined in (21) is monotonically increasing.*

**Proof.** According to equations (20) and (21), the following equation can be obtained

$$\tau(t) = -\frac{t}{\ln \lambda_\Sigma(t)}.\tag{22}$$

For $\forall t_1, t_2 \in [0, +\infty)$, let say, $t_1 < t_2$, then

$$\tau(t_2) - \tau(t_1) = -\frac{t_2}{\ln \lambda_\Sigma(t_2)} + \frac{t_1}{\ln \lambda_\Sigma(t_1)} = \frac{C(t)}{\ln \lambda_\Sigma(t_1) \ln \lambda_\Sigma(t_2)},\tag{23}$$

where

$$C(t) = -t_2 \ln \lambda_\Sigma(t_1) - (-t_1 \ln \lambda_\Sigma(t_2)).\tag{24}$$

Since it is easy to prove that $-\ln \lambda_\Sigma(t)$ is a monotonically decreasing function, we can get that

$$-\ln \lambda_\Sigma(t_1) > -\ln \lambda_\Sigma(t_2).\tag{25}$$

Thus

$$C(t) > 0.\tag{26}$$

Note that because the dominator of equation (23) is always positive, $\tau(t_2)$ is always greater than $\tau(t_1)$. Hence the time coefficient function $\tau(t)$ that satisfies equation (21) is monotonically increasing. This ends the proof. □

**Proposition 3.** *The values of the two boundaries of the scale factor function $\tau(t)$ are*

$$\begin{cases} \dfrac{1}{\tau_0} = \sum_{i=1}^{n} \rho_i \dfrac{1}{\tau_i} \\ \tau(+\infty) = \max\{\tau_1, \tau_2, ..., \tau_n\} \end{cases},\tag{27}$$

where

$$\tau_0 = \tau(0).\tag{28}$$

**Proof.** When $t = 0$, equation (22) is indefinite. Therefore, we define $\tau(0)$ via the L'Hopital's rule

$$\tau(0) = \tau_0 = -\lim_{t \to 0} \frac{\lambda_\Sigma(t)}{\dot{\lambda}_\Sigma(t)} = (\sum_{i=1}^{n} \rho_i \frac{1}{\tau_i})^{-1}.\tag{29}$$

Hence the first expression of equation (27) holds.

When $t \to +\infty$, equation (22) is also indefinite. Similarly, we define $\tau(+\infty)$ via the L'Hopital's rule

$$\tau(+\infty) = -\lim_{t \to +\infty} \frac{\lambda_\Sigma(t)}{\dot{\lambda}_\Sigma(t)}.\tag{30}$$

Note that the greater $\tau$ is, the slower the term $e^{-\frac{t}{\tau}}$ converges to zero. If

$$\hat{\tau} = \max\{\tau_1, \tau_2, ..., \tau_n\},\tag{31}$$

then

$$\begin{cases} \lim_{t \to +\infty} \lambda_\Sigma(t) = \lim_{t \to +\infty} \hat{\rho} e^{-\frac{t}{\hat{\tau}}} \\ \lim_{t \to +\infty} \dot{\lambda}_\Sigma(t) = -\lim_{t \to +\infty} \frac{\hat{\rho}}{\hat{\tau}} e^{-\frac{t}{\hat{\tau}}} \end{cases}, \tag{32}$$

where $\hat{\tau}$ represents the maximum among given $n$ values of $\tau$.

Thus

$$\tau(+\infty) = \lim_{t \to +\infty} \frac{\hat{\rho} e^{-\frac{t}{\hat{\tau}}}}{\frac{\hat{\rho}}{\hat{\tau}} e^{-\frac{t}{\hat{\tau}}}} = \hat{\tau}. \tag{33}$$

The second expression of equation (27) holds. This ends the proof. □

To better demonstrate the properties established above, we depict a possible group of curves in Figure 1 as an example. In this example, we assume there are two NSFFE ($\lambda(t)$ function), say $\lambda_1(t)$ and $\lambda_2(t)$. Their time coefficients are set as $\tau_1 = 4$, $\tau_2 = 2$ and their amplitudes are set as 0.8 and 0.2, respectively. Thus, the NSSF ($\lambda_\Sigma(t)$ function) can be expressed as

$$\lambda_\Sigma(t) = \lambda(t, \tau_1, \tau_2) = 0.8\lambda(t, 4) + 0.2\lambda(t, 2). \tag{34}$$

The initial value of $\tau(t)$ is

$$\tau(0) = \tau_0 = (\frac{0.8}{4} + \frac{0.2}{2})^{-1} = 3.33. \tag{35}$$

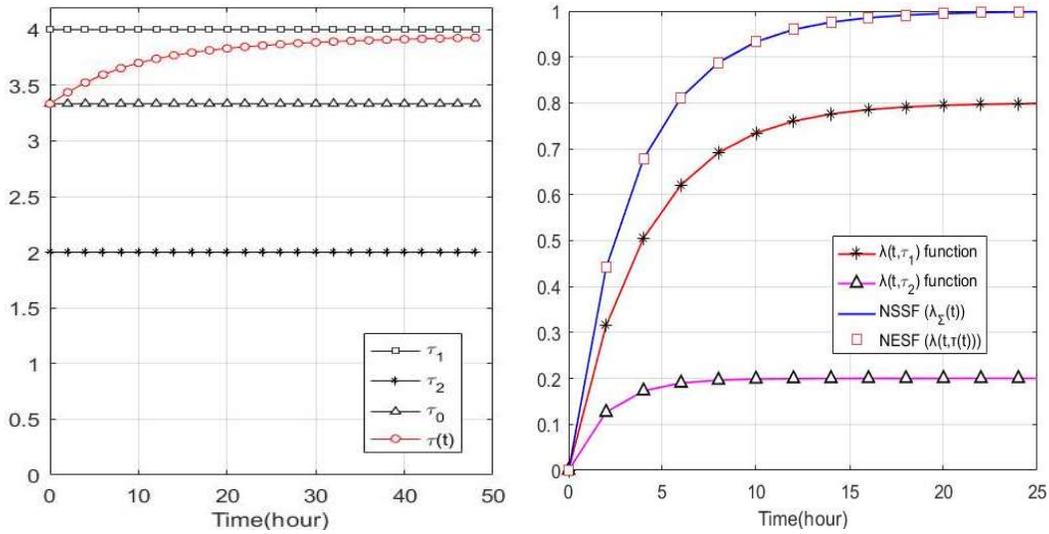

(a) Examples of $\tau(t)$ function.　　　　　　(b) Examples of $\lambda$ function.

**Figure 1.** Examples of $\tau(t)$ and $\lambda$ functions.

The curves in Figure 1(a) show that in this example, the time coefficient function $\tau(t)$ is indeed monotonically increasing. The initial value $\tau(0)$ is equal to $\tau_0 = 3.33$ and $\tau(+\infty)$ converges to $\tau_1 = 4$. The curves in Figure 1(b) indicate that the curve of the NESF ($\lambda(t, \tau(t))$ function), matches the curve of the NSSF ($\lambda_\Sigma(t)$ function). Hence these observations are consistent with the propositions established in this subsection.

*2.3 Contour Statistical Functions*

Given a group of SFFE ($\alpha(t)$) with $\{\tau_1, \tau_2, ..., \tau_n\}$, we define the Contour Statistical Function (CSF) with the time coefficient $\tau_0$ as an upper bound for the SSF

$$\alpha(t, \tau_0) \triangleq A_\Sigma(1 - e^{-\frac{t}{\tau_0}}). \tag{36}$$

The corresponding $\lambda$ function is referred to as the Normalized Contour Statistical Function (NCSF), that is

$$\lambda(t, \tau_0) \triangleq 1 - e^{-\frac{t}{\tau_0}}.$$ (37)

**Proposition 4.** *The NCSF ( $\lambda$ function) is greater than or equal to the NESF ( $\lambda$ function) and the maximum deviation point $t^*$ satisfies the following equation*

$$\frac{1}{\tau_\Sigma} e^{-\frac{t^*}{\tau_0}} = \sum_{i=1}^{n} \frac{\rho_i}{\tau_i} e^{-\frac{t^*}{\tau_i}}.$$ (38)

**Proof.** The time coefficient of the NCSF ( $\lambda(t, \tau_0)$ function) is the constant $\tau_0$, the time coefficient function $\tau(t)$ of equivalent error function is monotonically increasing, and the initial value is $\tau_0$. Therefore,

$$\Delta\lambda(t) = \lambda(t, \tau_0) - \lambda(t, \tau(t)) \ge 0,$$ (39)

that is, the NCSF is not less than the NESF.

Since $\Delta\lambda(0) = 0$ and $\Delta\lambda(+\infty) = 0$, the maximum deviation will not occur at the boundary.

Since equation (21) holds, then

$$\Delta\lambda(t) = \sum_{i=1}^{n} \rho_i e^{-\frac{t}{\tau_i}} - e^{-\frac{t}{\tau_0}}.$$ (40)

Because $\lambda_\Sigma(t)$ and $\lambda(t, \tau(t))$ are both monotonic, $\Delta\lambda(t)$ only has one maximum value. Thus, the corresponding zero derivative equation is

$$\Delta\lambda^{(1)}(t) = \frac{1}{\tau_\Sigma} e^{-\frac{t}{\tau_0}} - \sum_{i=1}^{n} \frac{\rho_i}{\tau_i} e^{-\frac{t}{\tau_i}} = 0.$$ (41)

Therefore, equation (38) holds. This ends the proof. □

**Corollary 2.** The inequality of the upper bound of the SSF ( $\alpha_\Sigma$ function), satisfies

$$\delta_\Sigma(t) \le \alpha(t, \tau_0).$$ (42)

**Proof of Corollary 2.** The corollary is self-evident. This ends the proof. □

Since $\Delta\lambda(t)$ can be transformed into

$$\Delta\lambda(t) = e^{-\frac{t}{\tau_\Sigma}} (\sum_{i=1}^{n} \rho_i e^{-\Delta(\frac{1}{\tau})_i t} - 1),$$ (43)

where

$$\Delta(\frac{1}{\tau})_i = \frac{1}{\tau_i} - \frac{1}{\tau_0}.$$ (44)

This indicates that the difference between the NESF and the NCSF, that is $\Delta\lambda(t)$, is related to the difference between the reciprocal of the time coefficients. A smaller differential value can be more helpful in reducing $\Delta\lambda(t)$.

The CSF defined in this subsection provides an upper bound for the ESF. Specifically, the CSF is equal to the ESF at points of $t = 0$ and $t = +\infty$, respectively. At other points, the CSF lies strictly above the ESF.

## 3. Quantifying the Statistical Forecast Uncertainty of Wind or Solar Power Source

### 3.1 Statistics

Based on a large amount of on-site data, previous studies [25,26] have established the relationship between the root mean square error (RMSE) of wind and solar power source forecasts and the time advance of day-ahead scheduling and look-ahead dispatch. In this paper, such a relation is referred to as the $\alpha^{rmse}$ curve.

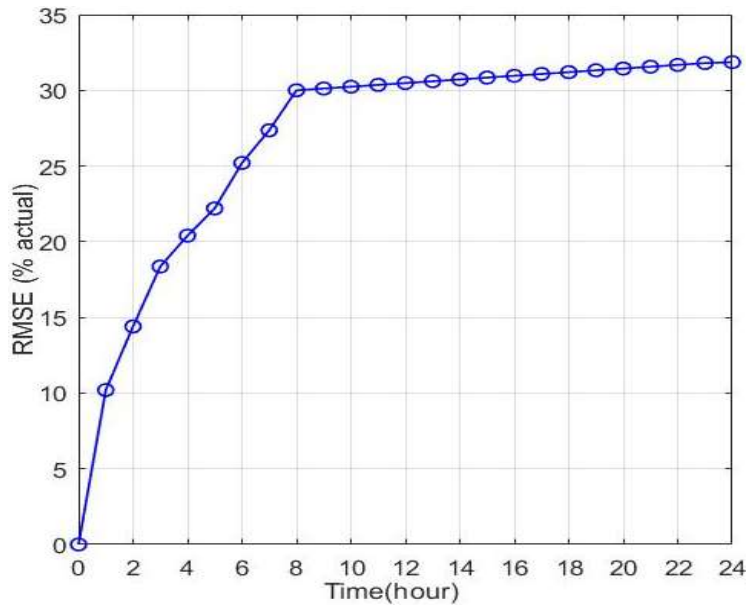

**Figure 2.** $\alpha_w^{rmse}$ curve.

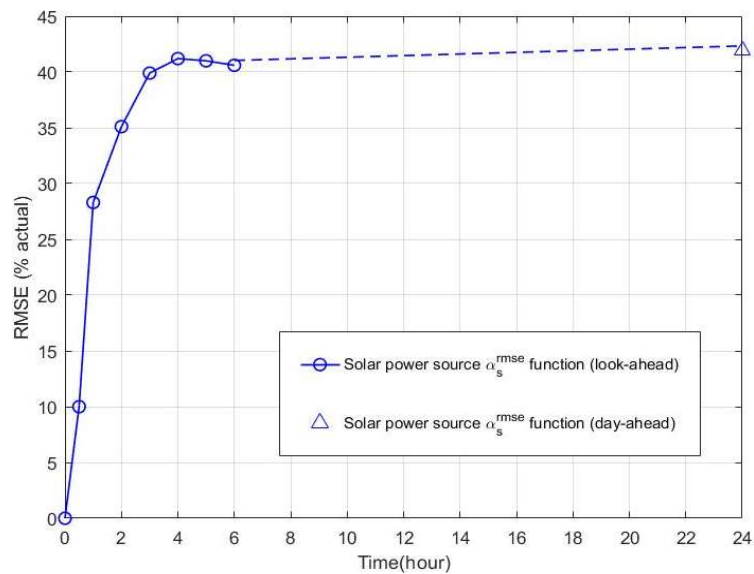

**Figure 3.** $\alpha_s^{rmse}$ curve.

Figure 2 shows the statistics of the $\alpha^{rmse}$ curve from a wind farm in the northwestern U.S., wherein each point represents the mean of RMSE of the forecast with given time advance (X-axis) over 3 years. The range of possible time advance spans from 24 hours to one hour with a step-size of one hour. Figure 3 exhibits a similar $\alpha^{rmse}$ curve from a photovoltaic power plant in northern France. The range of time advance spans from six hours to one hour. The "Δ" marker in Figure 3 represents the RMSE of the 24 hours-ahead forecast. Note that the RMSE in this paper represents the percentage of the relative forecast error with respect to the forecast power values and the actual power values.

Such $\alpha^{rmse}$ curves represent the statistics of the change of forecast error with respect to time advance, which is a noisy random process. From Figure 2 and Figure 3 we can see that the $\alpha^{rmse}$ curves of both the wind farm and the photovoltaic power plant are similar in shape to the $\alpha$ function curve defined in equation (1). This is also consistent with the general rule of the prediction of future events. As the advanced forecast time shortens, the event in the future will gradually become clearer. In particular, if the advanced forecast time reduces to a certain extent, the clarity will increase rapidly.

By utilizing a large enough amount of historical data, the $\alpha^{rmse}$ curve eliminates the randomness of individual examples and reveals the objective statistics in the prediction of events in the future. Therefore, the aforementioned $\alpha$ function can be employed to describe the statistical function of the forecast error of power generations from IPS.

*3.2 Statistical Functions of Forecast Error of Wind or Solar Power Source*

For a specific wind farm or photovoltaic power plant, the point with time advance $t_i$ in the $\alpha^{rmse}$ curve is defined as follows

$$\alpha^{rmse}(t_i) \triangleq \sqrt{\frac{1}{m}\sum_{k=1}^{m}\varepsilon_k^2(t_i)} \,, \tag{45}$$

where the scalar $m$ is the number of forecasting samples. $\varepsilon_k(t_i)$ is the relative error in the power prediction of sample $k$ when the advanced time is $t_i$, and can be expressed as follows

$$\varepsilon_k(t_i) = \frac{P_f^k(t_i) - P_a^k(t_i)}{P_a^k(t_i)}, \tag{46}$$

where $P_f^k(t_i)$, $P_a^k(t_i)$ represent the forecast power value and the actual power value of sample $k$ when the advanced time is $t_i$, respectively.

The $\alpha^{rmse}$ curve is represented by time sequence comprised of $\alpha^{rmse}(t_i)$

$$\alpha^{rmse} \triangleq \left\{\alpha^{rmse}(t_i)\right\}, \ (i=0,1,2,...,24)\,, \tag{47}$$

where $\alpha^{rmse}(t_0)$ is set as zero.

Regarding to all possible $t_i$, we employ $\alpha$ function, that is, the SFFE of equation (1) to approximate the real $\alpha^{rmse}$ curve. We set the parameters A and $\tau$ such that the resultant $\alpha$ function yields the following conditions:

1. Amplitude condition:

$$A = \max\left\{\alpha^{rmse}(t_i)\right\}. \tag{48}$$

2. Time coefficient condition:

$$\tau = \frac{A}{\alpha^{(1)}(0)} = \frac{A}{\max\dfrac{\alpha^{rmse}(t_{i+1}) - \alpha^{rmse}(t_i)}{t_{i+1} - t_i}}\,. \tag{49}$$

These conditions dedicate to customize the SFFE ($\alpha$ function), as a profile of the $\alpha^{rmse}$ curve. Although there is no rigorous proof, our intuition suggests that such a $\alpha$ function can include all points in the $\alpha^{rmse}$ curve underneath, which will be further demonstrated in Section 5. Hereafter, such a customized $\alpha$ function will be served as a mathematical model for the description of quantifying the uncertainties of power generations from wind or solar power source.

More specifically, we define the NSFFE ($\lambda$ function) of wind power source as

$$\lambda(t,\tau_w) = 1 - e^{-\frac{t}{\tau_w}}\,, \tag{50}$$

And define the corresponding SFFE ($\alpha$ function) as

$$\alpha_w(t) = A_w\lambda(t,\tau_w)\,. \tag{51}$$

where $A_w$ is represented as the amplitude of SFFE of wind power source.

Similarly, we define the NSFFE ($\lambda$ function) of solar power source as

$$\lambda(t,\tau_s) = 1 - e^{-\frac{t}{\tau_s}}\,, \tag{52}$$

and the corresponding SFFE ($\alpha$ function) is defined as

$$\alpha_s(t) = A_s\lambda(t,\tau_s)\,. \tag{53}$$

where $A_s$ is represented as the amplitude of SFFE of solar power source.

For wind and solar power sources, the time coefficients of each are expressed in terms of $\tau_w$ and $\tau_s$, respectively. Their values can be obtained by equation (49).

In practice, values of $\tau_w$ and $\tau_s$ are usually far less than 24 hours. There are only minor differences between the day-ahead uncertainty

$$\alpha_{w24} = \alpha_w(24h), \ \alpha_{s24} = \alpha_s(24h) \tag{54}$$

and the amplitude of the SFFE. Therefore, the amplitudes can be approximated as

$$A_w = \alpha_{w24}, \ A_s = \alpha_{s24}. \tag{55}$$

## 4. Quantifying the Statistical Forecast Uncertainty of the Power System Sources

*4.1 Statistical Forecast Uncertainty of IPS*

### 4.1.1. Sum Statistical Functions of IPS

Without loss of generality, this paper assumes that the IPS consist of wind and solar power sources. Let $P_w$ and $P_s$ be, respectively, the power generation of the wind and solar power sources, then we merge them into the overall power generations from IPS, let say $P_{ips}$, that is

$$P_{ips}(t) = P_w(t) + P_s(t). \tag{56}$$

The proportion of the power generation $P_w$ of wind farms in the overall IPS power generations $P_{ips}$ is represented by $\beta_w$, which is defined as

$$\beta_w(t) \triangleq \frac{P_w(t)}{P_{ips}(t)}, \ (\beta_w(t) \in [0,1]). \tag{57}$$

The proportion of the solar power is

$$\frac{P_s(t)}{P_{ips}(t)} = 1 - \beta_w(t). \tag{58}$$

As described in Section 2, here we define $\alpha_{ips}(t)$ as the SSF of IPS (which is $\alpha_\Sigma(t)$) to represent the overall statistical uncertainty of IPS. Consequently, equation (12) becomes

$$\alpha_{ips}(t) = \beta_w(t)\alpha_w(t) + (1-\beta_w(t))\alpha_s(t). \tag{59}$$

In this paper, $\beta_w(t)$ is an actual proportion value, which represents the proportion of the wind power generation in IPS generation when time advance is zero. Here we assume $\beta_w(t) = \beta_w$, then the SSF of IPS ($\alpha_{ips}(t)$ function) can be simplified as

$$\alpha_{ips}(t) = \beta_w\alpha_w(t) + (1-\beta_w)\alpha_s(t). \tag{60}$$

The equation above is equivalent to

$$\alpha_{ips}(t) = \beta_w A_w\lambda(t,\tau_w) + (1-\beta_w) A_s\lambda(t,\tau_s), \tag{61}$$

where

$$A_{ips} = \beta_w A_w + (1-\beta_w) A_s. \tag{62}$$

We further define $\gamma$ coefficient as the following

$$\gamma \triangleq \beta_w \frac{A_w}{A_{ips}}, \tag{63}$$

then the NSSF of IPS ($\lambda_{ips}(t)$ function) can be written as

$$\lambda_{ips}(t) = \gamma\lambda(t,\tau_w) + (1-\gamma)\lambda(t,\tau_s). \tag{64}$$

### 4.1.2 Equivalent Statistical Functions of IPS

The ESF that is equivalent to the SSF ($\alpha_{ips}(t)$ function) is as follows

$$\alpha(t,\tau(t)) = \alpha_{ips}(t) = A_{ips}(1-e^{-\frac{t}{\tau(t)}}), \tag{65}$$

where $A_{ips}$ denotes the amplitude of SFFE of IPS.

The NESF that corresponds to the ESF ($\alpha(t,\tau(t))$ function) is shown as follows

$$\lambda(t, \tau(t)) = \lambda_{ips}(t) = 1 - e^{-\frac{t}{\tau(t)}} . \tag{66}$$

### 4.1.3 Contour Statistical Functions of IPS

Furthermore, we employ the CSF of IPS to establish a closure of the ESF. The CSF ( $\alpha(t, \tau_0)$ function) is shown as

$$\alpha(t, \tau_0) = A_{ips}(1 - e^{-\frac{t}{\tau_0}}) , \tag{67}$$

where

$$\tau_0 = \tau(0) = (\frac{\gamma}{\tau_w} + \frac{1 - \gamma}{\tau_s})^{-1} . \tag{68}$$

The NCSF ( $\lambda(t, \tau_0)$ function) is expressed as

$$\lambda_{ips}(t, \tau_0) = \frac{1}{A_{ips}} \alpha_{ips}(t, \tau_0) . \tag{69}$$

### 4.2 Statistical Forecast Uncertainty of All Power Sources

Besides IPS, the power system is also comprised of controllable generation assets such as thermal power, hydropower, nuclear power and energy storage facilities. Let the power generation of the controllable generation assets be $P_c$, then the overall power generation from the power system, let say $P_g$, can be expressed as follows

$$P_g(t) = P_c(t) + P_{ips}(t) . \tag{70}$$

Herein, the IPS power proportion $P_{ips}(t)$ relative to $P_g(t)$ is referred to as the $\beta_{ips}$ proportion, that is

$$\beta_{ips}(t) \triangleq \frac{P_{ips}(t)}{P_g(t)}, \ (\beta_{ips}(t) \in [0,1]) . \tag{71}$$

In this paper, $\beta_{ips}(t)$ is an actual proportion value, which represents the proportion of the IPS power generation in the overall power generation when time advance is zero. Here we assume that $\beta_{ips}(t)$ is a constant $\beta_{ips}$. We define $\alpha_g(t)$ as the SSF of all power sources (which is $\alpha_\Sigma$) to represent the overall statistical uncertainty of all power sources. Hence, we get that

$$\alpha_g(t) = \beta_{ips} \alpha_{ips}(t) . \tag{72}$$

The ESF of all power sources ( $\alpha_g(t, \tau(t))$ function) is

$$\alpha_g(t, \tau(t)) = A_g(1 - e^{-\frac{t}{\tau(t)}}) = A_g \lambda_g(t, \tau(t)) , \tag{73}$$

where $A_g$ is represented as the amplitude of SFFE of all power sources, and it satisfies

$$A_g = \beta_{ips} A_{ips} = \beta_{ips} \beta_w A_w + (\beta_{ips} - \beta_{ips} \beta_w) A_s . \tag{74}$$

Note that in equation (73), the NESF ( $\lambda_g(t, \tau(t))$ function) that corresponds to SSF of all power sources $\alpha_g(t, \tau(t))$ is the same as that of the IPS. This indicates that the $\beta_{ips}$ proportion has nothing to do with the time coefficient function, that is, the time coefficient function exhibits "inheritance". In equation (74), the amplitude $A_g$ of the SSF of all power sources is $\beta_{ips}$ times less than the amplitude $A_{ips}$ of the IPS, which exhibits "dilution".

The CSF of all power sources is

$$\alpha_g(t, \tau_0) = A_g \lambda(t, \tau_0) . \tag{75}$$

From the perspective of power systems, the SFFE of wind or solar power source ( $\alpha_w$ function or $\alpha_s$ function) and the SFFE of IPS ( $\alpha_{ips}$ function) will be ultimately merged as the SFFE of all power sources ( $\alpha_g$ function) which quantifies the statistical uncertainty in the overall power generation. This function provides an important reference for many power system applications such as the real-time power balancing, security analysis, and electric power quality control.

## 5. Examples

### 5.1 Uncertainty of Wind or Solar Power Source

Taking the statistical forecast error curve presented in [25, 26] as an example, the SFFE of wind or solar power source ($\alpha$ function) can be respectively determined based on the curves of $\alpha_w^{rmse}$ and $\alpha_s^{rmse}$ shown in Figure 2 and Figure 3. According to equations (48) and (49), the resulting SFFE can be obtained as follows

$$\begin{cases} \alpha_w(t) = 31.86 \times (1 - e^{-\frac{t}{2.67}}) \ (\%) \\ \alpha_s(t) = 41.90 \times (1 - e^{-\frac{t}{0.89}}) \ (\%) \end{cases}. \tag{76}$$

In this example, the amplitude constraint and the derivative constraint are both satisfied. As a result, the resulting SFFE ($\alpha_w(t)$ function and $\alpha_s(t)$ function) define reasonable closures for the real $\alpha_w^{rmse}$ and $\alpha_s^{rmse}$ curves, respectively, as shown in Figure 4. In other words, the $\alpha_w$ and $\alpha_s$ functions respectively "adhere" to the sequences $\alpha_w^{rmse}$ and $\alpha_s^{rmse}$ "from outside toward inside"; this shows that the $\alpha$ function can well describe the statistical uncertainty of the wind or solar power source.

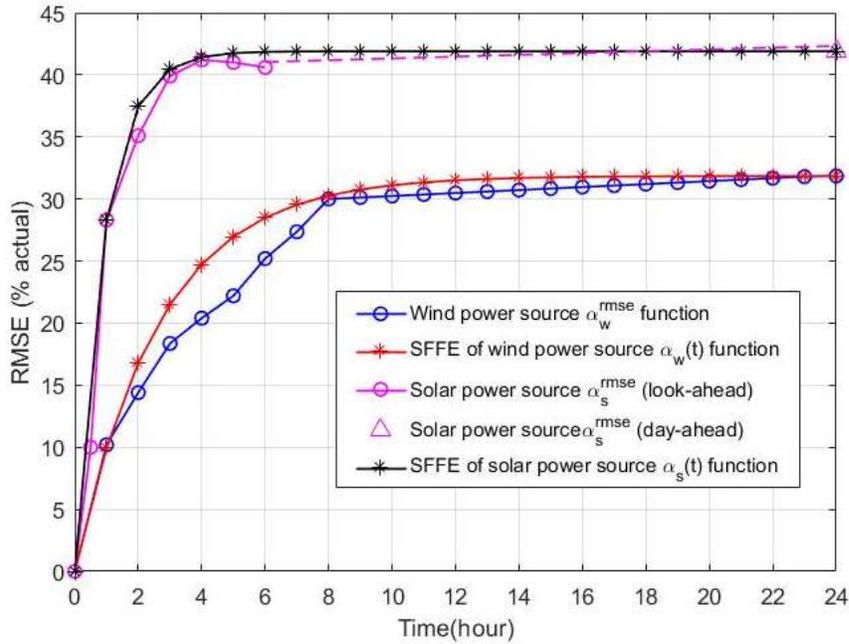

**Figure 4.** $\delta^{rmse}$ sequences and $\delta$ functions of wind and solar power sources.

### 5.2 Uncertainty of IPS

For a power system with IPS, the SFFE of wind and solar power sources, $\alpha_w$ function and $\alpha_s$ function, are shown in equation (76). Set $\beta_w$ proportion at 80%. From equation (60), the SSF of IPS, $\alpha_{ips}(t)$ function, is expressed as

$$\alpha_{ips}(t) = 22.49 \times (1 - e^{-\frac{t}{2.67}}) + 8.38 \times (1 - e^{-\frac{t}{0.89}}), \tag{77}$$

and

$$A_{ips} = 0.8A_w + 0.2A_s = 33.87\% . \tag{78}$$

The $\gamma$ coefficient is

$$\gamma = \beta_w \frac{A_w}{A_{ips}} = 0.75 . \tag{79}$$

The time coefficient $\tau_0$ of CSF of IPS can be obtained as

$$\tau_0 = (\frac{\gamma}{\tau_w} + \frac{1-\gamma}{\tau_s})^{-1} = 1.79\,(h) \tag{80}$$

Therefore, the CSF of IPS is

$$\alpha_{ips}(t, \tau_0) = 33.87 \times (1 - e^{-\frac{t}{1.79}})\,(\%)\,. \tag{81}$$

Figure 5 shows that the CSF of IPS "adheres" to the SSF of IPS "from outside toward inside", and the maximum deviation between the two is 2.24% (as denoted by "*" in the figure). This indicates that the CSF of IPS can well depict the statistical uncertainty of IPS.

### 5.3 Uncertainty of All Power Sources

For a power system consisting of IPS and controllable generators, the corresponding SSF of IPS is shown in equation (77). Set the $\beta_{ips}$ proportion at 60%. Then, based on equation (74), the CSF of all power sources is

$$\alpha_g(t, \tau_0) = 20.32 \times (1 - e^{-\frac{t}{1.79}})\,\%\,. \tag{82}$$

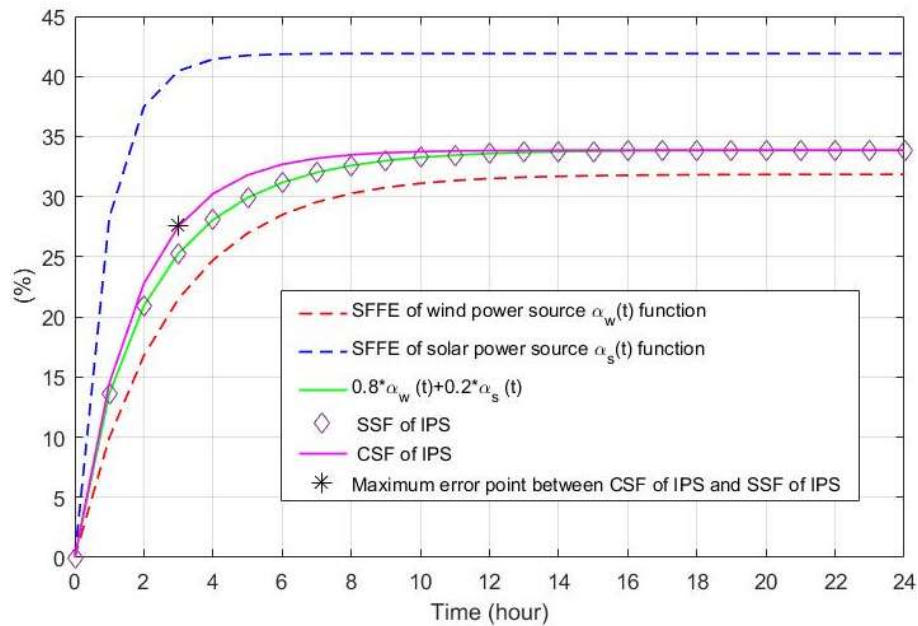

**Figure 5.** SSF and CSF of IPS.

## 6. Conclusions and Practical Applications

The conclusions of this paper include the following:

1. A negative-exponential forecast uncertainty function is proposed to statistically quantify the uncertainty of a single IPS. Several basic properties of the statistical function have been established.

2. To statistically quantify the uncertainty of multiple IPS as well as all power sources, a sum statistical function is defined as the sum of multiple forecast uncertainty functions, which can also be represented with a single equivalent statistical function with variant time coefficient. A contour statistical function with a constant time coefficient is also designed to provide a closure of the equivalent statistical function.

3. Historical observations of forecast errors match well with the proposed statistical functions in temporal properties. This is not an incidental phenomenon, but reflects the general statistical law of the prediction of future events.

4. The proposed statistical functions of forecast error are leveraged to quantify the statistical uncertainty level of a single IPS, multiple IPS, as well as all power sources, whose parameters are estimated based on historical data. The final contour statistical function is a succinct representation of the statistical uncertainty level of all power sources in the power system.

In summary, the contour statistical function represents the overall statistical uncertainty of all power sources. The amplitude and time coefficient are the two numerical characteristics that can well depict the statistical uncertainty.

The proposed statistical functions can be utilized in many important practical applications as detailed below:

First, the comprehensive assessment of the IPS quality can be better achieved by considering statistical functions with multiple factors like wind speed, wind direction, irradiation time and irradiation intensity.

Second, from the perspective of the real-time power balancing, the proposed statistical functions can determine the factors of the timescale of the real-time dispatch with high proportion of IPS generation. Combined with the power balancing capability of Automatic Generation Control (AGC), an analytical formulation about the critical real-time dispatch timescale can be further achieved. This study has been accomplished by the authors of this paper already.

Third, the proposed statistical functions can provide more quantitative characterization towards the statistics of the intermittent power uncertainty corresponding to different methodologies of power network planning and operation under high proportion of IPS generation, like power system security analysis, reliability assessment, and economical operation.